\documentclass[conference]{IEEEtran}
\IEEEoverridecommandlockouts
\usepackage{cite}
\usepackage{amsmath,amssymb,amsfonts}
\usepackage{algorithmic}
\usepackage{graphicx}
\usepackage{textcomp}
\usepackage{xcolor}
\usepackage{array, multirow}
\usepackage{caption}
\usepackage{subcaption}
\usepackage{float}
\def\BibTeX{{\rm B\kern-.05em{\sc i\kern-.025em b}\kern-.08em
    T\kern-.1667em\lower.7ex\hbox{E}\kern-.125emX}}
\begin{document}

\title{A Novel Dimension Reduction Scheme for Intrusion Detection Systems in IoT Environments}

	\author{\IEEEauthorblockN{Amir Andalib }
	\IEEEauthorblockA{\textit{School of Electrical Engineering} \\
		\textit{Iran University of Science and Technology}\\
		Tehran, Iran \\
		amir\_andalib@elec.iust.ac.ir}
	\and
	\IEEEauthorblockN{Vahid Tabataba Vakili}
	\IEEEauthorblockA{\textit{School of Electrical Engineering  } \\
		\textit{Iran University of Science and Technology}\\
		Tehran, Iran \\
		vakily@iust.ac.ir}
}
%

\maketitle

\begin{abstract}
Internet of Things (IoT) brings new challenges to the security solutions of computer networks. So far, intrusion detection system (IDS) is one of the effective security tools, but the vast amount of data that is generated by heterogeneous protocols and "things" alongside the constrained resources of the hosts, make some of the present IDS schemes defeated. To grant IDSs the ability of working in the IoT environments, in this paper, we propose a new distributed dimension reduction scheme which addresses the limited resources challenge.  A novel autoencoder (AE) designed, and it learns to generate a latent space. Then, the constrained hosts/probes use the  generated weights to lower the dimension with a single operation. The compressed data is transferred to a central IDS server to verify the traffic type. This scheme aims to lower the needed bandwidth to transfer data by compressing it and also reduce the overhead of the compression task in the hosts. The proposed scheme is evaluated on three well-known network traffic datasets (UNSW-NB15, TON\_IoT20 and NSL-KDD), and the results show that we can have a 3-dimensional latent space (about 90\% compression) without any remarkable fall in IDS detection accuracy. 
\end{abstract}


\section{Introduction}
%
%
Nowadays, Internet of Things (IoT) is one of the intriguing buzzwords. It is defined as an interconnection of smart devices via the Internet. And the intelligence which is added to everyday devices and tools by embedded components that enables computing and communication abilities defines smart devices. IoT has found its way to a wide range of applications such as smart homes, automated industrial environments, healthcare, smart grid, agriculture, etc. \cite{survey10}. The growth in the number of connected smart devices surpasses the projected numbers, with regard to forbes.com
\footnote{https://www.forbes.com/sites/louiscolumbus/2016/11/27/roundup-of-internet-of-things-forecasts-and-market-estimates-2016/\#7989fd89292d} reports,
in 2025 there will be 75 billion connected devices in IoT.

However the integration of services and smart device can literally provide the access of anything from anywhere, this amount of interconnected devices alongside the sensitivity and importance of the information that is carried out by IoT networks signifies the importance of its security. 
A set of practices that aimed at providing security to computer networks, users and data is called cybersecurity. It includes secured protocols, log audit tools, firewalls, etc. One of vital and promising components of cybersecurity is intrusion detection system (IDS). Due to its many advantages, more importantly its robustness and potential of behaving intelligently, it attracts a lot of research attention since its announced in late eighties \cite{survey3}. In a comprehensive survey \cite{survey11} by N.Moustafa et al. IDSs are classified into four main types from the operational standpoint: 1. Misuse (Singnature)-based, 2. Anomaly-based, 3. Stateful protocol analysis-based, 4. Hybrid-based. 

Although IDS has the merit of robustness and immediate response to zero-day attacks in traditional networks, because of some intrinsic differences of the internet of thing (IoT) structure previous IDS schemes cannot be applied directly into the IoT environments. There are some barriers in the way of implementing IDSs in the IoT, such as new protocols, constrained computational and storage resources, heterogeneous devices, and a vast amount of data generated by nodes\cite{survey14, survey11}.
To address the constrained resources and bandwidth overhead challenges an asymmetric deep autoencoder(AE) is proposed in this paper to compress the network traffic data with  low computational resources needed at probes or IoT end nodes.
The compressed data can be transferred to the server via the network or isolated platforms like USB interface. Because of compression, transferring data use a very low segment of network bandwidth. On the server end, we use a random forest (RF) as a decision engine to decide about traffic type, whether it is normal or attack. 
We evaluate our proposed compression framework on three different network traffic datasets: UNSW-NB15, NSL-KDD and TON\_IoT20. The rest of this paper is organized as follows. Section \ref{background} provides a     brief background on in-use learning techniques and datasets. In Section \ref{relatedworks}, we give an account of related researches with a focus on dimension reduction. In Section \ref{systemarchi}, our proposed dimension reduction scheme is discussed. Experimental results are detailed in Section \ref{experimental}. Finally, Section \ref{conclusion} concludes the paper. 

\section{Background}
\label{background}
This section provides some primitive knowledge about: 1. In-use machine learning techniques, 2. Evaluation datasets.
\subsection{Autoencoders}
\begin{figure}
	\centering
	\includegraphics[scale=0.4]{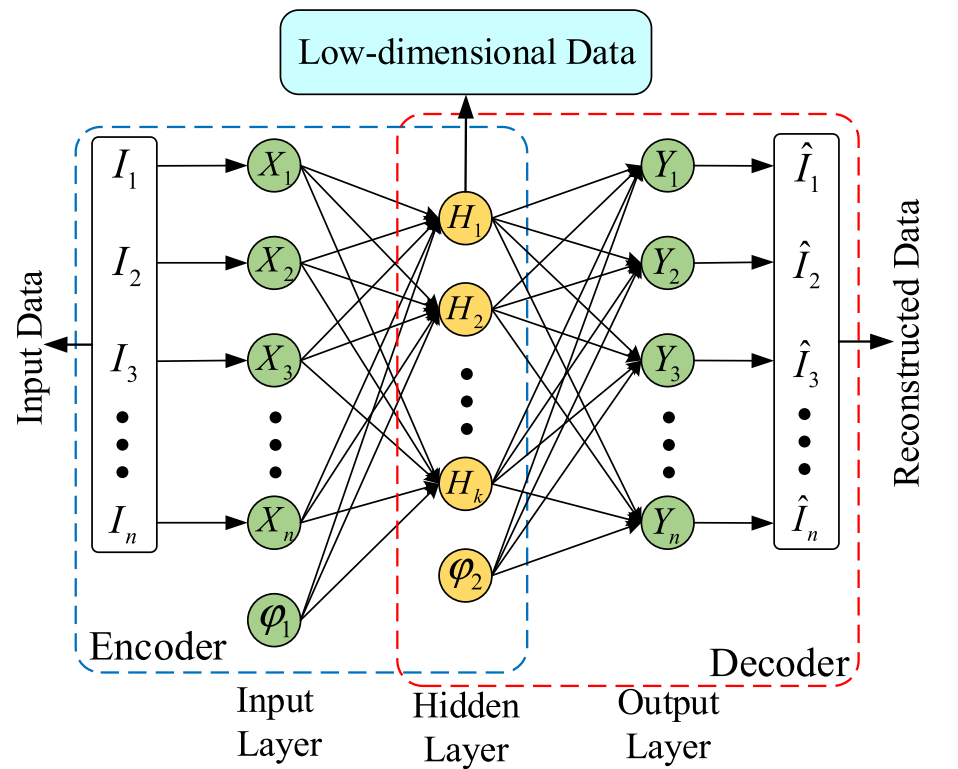}
	\caption{A simple Autoencoder \cite {ssae}}
	\label {simple_ae}
\end{figure}

Autoencoder (AE) is a multi-layer neural network composed of two stages: encoder and decoder. The goal of the encoder stage is to learn worthwhile properties of data and represent it in a lower dimension which is called latent space(LS).  Figure \ref{simple_ae} shows a schematic simple autoencoder. The goal of the decoder is to reproduce the original data by using the new representation provided by latent space. 
For a one-layer encoder, the latent space can be calculated by \cite{ssae}:
\begin{equation}
\mathbf{Z} =  \sigma(\mathbf{W_{x2z}X}+\mathbf{\phi_0}) 
\label{encoder_eq},
\end{equation}
where, $\mathbf{Z}‌$ is the latent space vector, $\mathbf{W_{x2z}}$, is the weight matrix to change the dimension of input ($\mathbf{X}$) , $\mathbf{\phi_0 }$ is the bias vector and $\sigma$ is the sigmoid function:
\begin{equation}
\sigma(x) = \frac{1}{1+e^{-x}}
\label{sigmoid_eq}.
\end{equation}
If the network succeeds in obtaining a close enough (closeness is defined by an error threshold) outputs to the inputs, then, the encoded version of data is a competent representative/surrogate. Generally AE is an unsupervised technique and beside compression, it is been applied in various other applications such as anomaly detection and handling missing values of data.

\subsection{Random Forest}
Random Forest(RF) is an ensemble learning technique which uses a large number of specialized decision trees (DTs)\cite{chollet}. DTs has tree-like structures made up of nodes and branches. In every node a feature is picked up to form the upcoming branch. DTs use mutual information (Information Gain) to select the best attribute based on which a node is formed. Although this method helps to find the best attribute to divide the data set in the immediate next step, it can not say which attribute will do better in several steps further. For example, choosing an attribute that does not have the best information gain for the present step, may lead to a better overall result in the next 5 steps. Furthermore, DTs suffer from overfitting to the training set. RFs resolve the aforementioned issues by randomly dividing the training set into several sets and using  different decision tree for each of them. Then, the outcomes of tree 	are combined to announce final result\cite{goodfellow}.  By utilizing the Law of Large Numbers, in \cite{rf}, L.Breimann gives a theoretical background for RFs; showing that they always converge and overfitting is not a problem anymore.

\subsection{Evaluation Datasets}
Commonly, the decision center of an anomaly-based IDS is implemented by machine learning techniques. Machine learning approaches need datasets to learn patterns. Because of the complexity and sophistication in computer network traffics, the validity and genuineness of the dataset are vital for the  IDS evaluation. Some reputed universities and institutions provide reliable datasets. In this work, we evaluate our proposed IDS on three of such sets. One of them is NSL-KDD introduced by Tavallee and et al. in \cite{nsl}(2004) as a purified version of the KDD99 dataset for the KDD cup (International Knowledge Discovery and Data Mining Tools Competition). It is an old and insufficient dataset but since it is been the main evaluation dataset for many field research papers, it is a sound reference to compare performances. 

As an up-to-date set with more sophisticated network traffic and attacks, we use UNSW-NB15 dataset. It was generated by Cyber Range Lab of UNSW \footnote{University of New South Wales}, and it consists of modern real-world network traffic and contemporary synthesized malicious activities.  About 100 GB of network packets are captured and analyzed  by Argus and IDS-Bro tools. 

Another modern dataset is TON\_IoT(UNSW-IoT20) dataset. It is generated by the IoT lab of the UNSW's electrical engineering school. It includes heterogeneous data collected from sources such as IoT sensors, windows and ubuntu operating systems\cite{ton_iot20}. The Table \ref{brief_config} presents a brief details about each of datasets and comparison between them.

\begin{table}
	\tiny
	\caption {Summary of Proposed System Configuration}
	\label{brief_config}
	\begin{center}
		\begin{tabular}{|l|c|c|c|}
			\hline
			Parameters					&	NSL-KDD			&	UNSW-NB15		&  TON\_IoT20	\\	\hline
			Total no. of records		&	148516			&	2540047			&  461043		\\ 	\hline
			No. of normal records		&	77053(~52\%)	&	1797198(~87\%)	&  300000(~65\%)\\	\hline
			No. of attack records		&	71463(~48\%)	&   260239(~13\%)	&  161043(35\%)	\\	\hline
			Attack Types				&	4 				&	   9			&   9\\	\hline
			No. of extracted features	&	41				&	   47           & 44		\\	\hline
			IoT Protocols	 			&	No				&	No				&Yes  			\\	\hline
			
			\hline
		\end{tabular}
	\end{center}
\end{table}

\section{Related works}
\label{relatedworks}
In this section, we provide some state-of-the-art  IoT intrusion detection systems with a special attention to the literature which consider dimension reduction, feature selection or feature mapping.


Recall that IoT systems use several new protocols in addition to the Internet such as IPv6 over low power wireless personal area network (6LoWPAN).  In one of the earliest papers that address the IoT IDSs (\cite{survey5r9}), Raza et al. propose an IDS named "SVELTE" and use both distributed and centralized deployment. It is made up of three components, first, a mapper  reconstructs RPL
\footnote{IPv6 Routing Protocol for Low-power and lossy networks} network traffic in the border router. Then, an intrusion detection module uses several algorithms such as network graph consistency  and node availability check to detect special types of routing attacks such as spoofing and sinkhole attacks. The other component is a mini-weight firewall in the nodes which aim to reduce the overhead of the central intrusion detection component by filtering pre-defined unwanted traffic. Their approach yields $100\%$ and $90\%$ true positive rate for sinkhole attack detection in lossless and lossy networks, respectively. 

Proper feature selection can result in lower computational overhead without remarkably impacting detection accuracy of the classifier. For instance, in \cite{ssae}, Yan et al. use sparse autoencoders (SAE) to compress data. They apply variety of layer combination in their SAE. The SAE follows by a simple classifier such as support vector machine (SVM), and random forest. The reasonable structure, training time-wise (~ $5 (sec)$), is a 5-layer encoder which yields overall 98.63\% of detection accuracy on NSL-KDD by using a latent space with five nodes.

In \cite{2tier_r39}, De La Hoz et al. use principle component analysis (PCA) and Fisher discriminant ratio (FDR) for feature selecting and denoising the data. For pattern recognition, they use probabilistic self-organizing map (PSOM). They evaluate their proposed system on the NSL-KDD dataset and by reducing the number of features to 20, the system achieve 93\% of detection accuracy.
\cite{tse} is another paper that uses a combination of feature selection techniques. Particle swarm optimization, colony, and genetic algorithms are used where a reduced error pruning tree classifier compares their results and chooses the best feature combination. Their feature selection procedure follows by a two-level classifier. They evaluate their proposed system on two datasets, NSL-KDD and UNSW-NB15 attaining 85.8 \% and 91.27\% detection accuracy, respectively.

Aside from computational complexity and training time reduction, lowering the dimension of data can have positive impacts on false positive detection. A two-layer dimension reduction technique  is introduced in \cite{2tier}. It shows a lower false alarm rate (FAR) with the same classifier when there is no dimension reduction (4.86\% vs. 5.44\% FAR). In this approach, the first layer uses PCA as an unsupervised method to reduce the dimension of NSL-KDD set from 41 to 35. The output of the second reduction layer is a 4D feature set which uses the linear discriminant analysis (LDA). After compression, k-nearest neighbor (KNN) and naive Bayes method are applied to data as classifiers which result in 84.86\% detection accuracy. 

For another dimension reduction technique a modified binary grey wolf optimization (GWO) is proposed by Alzubi et al. \cite{tse21}. GWO is a meta-heuristic optimization technique introduced by Mirjalali et al \cite{gwo}. The original algorithm is not suitable for dimension reduction, generally. The modified algorithm uses a supervised version of GWO and defines a fitness function to decide which subset of features to be chosen. The fitness function is:
\begin{equation}
\label{tse21_eq29}
Fitness = P.a + \frac{1}{NF}.b,
\end{equation}
where P, is the parameter that shows the goal of optimization (for instance, the detection accuracy of classifier), a and b are empirical coefficients, and $NF$ is the number of features in target reduced space.
In every iteration of the algorithm, a subset of features is chosen, then an SVM machine is learned to classify the samples. Based on the results of the classification and P, the fitness function value is calculated and compared to the threshold until the target reached. They evaluated their proposed algorithm on the NSL-KDD dataset. By running the algorithm for 20 times, a subset of features with the size of 26 out of 41 is selected by the algorithm and it yields 81.58\% detection accuracy.

\section {Proposed Methodology}
\label{systemarchi}
Constrained hosts and nodes in the IoT networks is one of the destructive barriers in the way of applying traditional IDS solutions directly into these networks. On the other hand, the amount and  variety of data that is been generated by IoT devices are very high which makes transferring all traffic to one central IDS server, an obsolete solution. It also renders IDS itself vulnerable to attacks such as DoS. To tackle these issues we propose a dimension reduction scheme using deep learning techniques and an IDS framework that can operate properly in IoT environments. In the following subsections, we discuss our proposed scheme in detail.

\subsection{Proposed Dimension Reduction Scheme}
The main goal of our work is to propose a dimension reduction scheme which has two main characteristics:
\begin{enumerate}
	\item{The competence of compressed data  to train the decision engine of an anomaly-based IDS}
	\item{The scheme ability to impose  low computational overhead to the low-end IoT devices.}
\end{enumerate}
To this end, we propose a novel asymmetric supervised autoencoder.  
Every layer in a deep neural network depicts a different representation of input data. Autoencoders use this attribute to give a compact representation of data that can be used to reproduce the original data with an acceptable amount of error. 
\subsection{Encoder} 	
\begin{figure}
	\centering
	\includegraphics[scale=0.6]{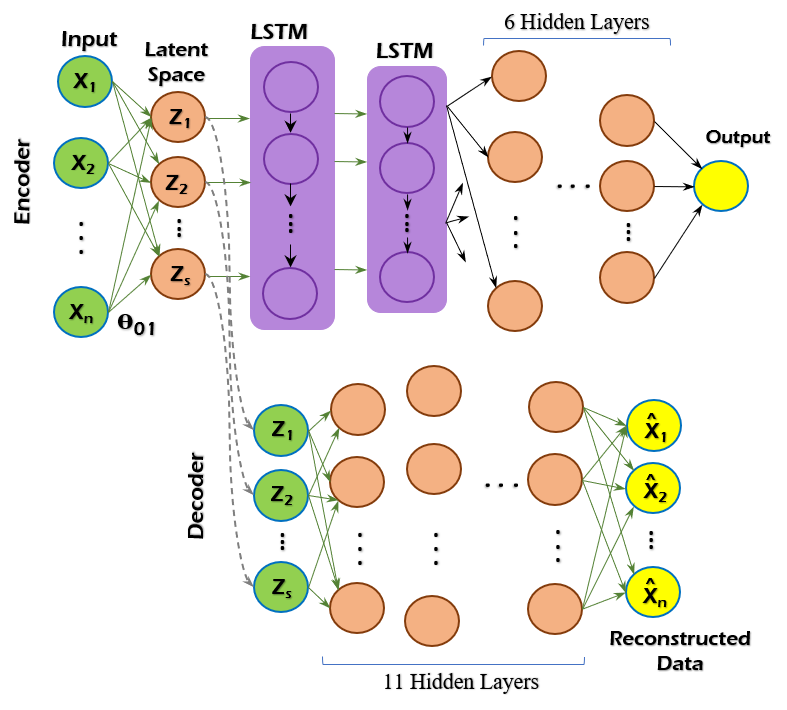}
	\caption{Proposed Autoencoder}
	\label {proposed_ae}
\end{figure}

As shown in Figure \ref{proposed_ae}, our proposed encoder is made up of a dense layer which its output is the encoded data, two LSTM layers, and an MLP consists of 6 hidden layers. Usually, autoencoders are made symmetric and the latent space is the output of the middle layer. Through an unsupervised procedure, a multi-layer network strives to regenerate the input data at the output end. But here we use a standalone classifier as an encoder. Like a neural network, it gets trained with a labeled dataset and every layer is a variant representation of the input data. The first hidden layer is chosen as the latent space. The obtained weight matrix that maps the input high dimension data to encoded data is deployed to constrained hosts. The hosts have to carry out a matrix multiplication and that is the only computation they need to do. This is the rationale behind embedding the latent space on the top, unlike conventional autoencoders. 

Since there is no rigorous theoretical background of different deep neural network configurations' functionality
\cite{deepmath}, usually the trial and error method is applied to designate the competent architecture. Hence, we try different neural networks and choose our encoder network based on two criteria: 1. The classifier accuracy in detection
\begin{equation}
\label{acc}
Acc.  = \frac {True \  Postitive + True \  Negative}{No. \ of \  All \  Samples},
\end{equation}
and 2. Mean square error of reconstruction (decoder):
\begin{equation}
\label{mse}
MSE = \frac{1}{N}\sum\limits_{i=1}^{N} (X_{input} - X_{Reconstructed})^2,
\end{equation}
where $N$ is the number of features in the original data space. The reconstruction error is considered to choose both the encoder and decoder configuration. The results show that whenever this network trained well (high accuracy), the output of every layer can be used to reconstruct data with a proper decoder. The proposed encoder as a classifier uses two layers of  LSTM units, these units are memory-based modules and they can choose what to remember and what to forget. The sequential nature of the network traffic data makes it imperative to perform a time series analysis of the data. Here, this analysis is done by recurrent layers. 
\subsection{Decoder}
For decoder, variety of combinations including an LSTM-based decoder are examined and we choose the network which is able to generate the original data with a minimum mean square error. Figure \ref{proposed_ae} shows the neural network that is been used as the decoder. It is a 11-layer MLP that roughly has the mirrored layer sequence of the encoder's MLP unit. Reconstructing the original data may have many usages, but our final goal of compression is to run an IDS with acceptable functionality, in which, the operation criteria like detection accuracy, precision, and false alarm rate remain as close as it is possible to the case of using original data. 

To find the proper configuration for both decoder and encoder a set of configurations based on a heuristic approach is defined and then bayesian optimization is used to obtain a suboptimum configuration. Similar to many deep learning approaches, simulation and validation are inseparable parts of system design.

\section {Experimental Results}
\label{experimental}
Our concluded architectures for encoder and decoder are based on validating them on UNSW-NB15 dataset. Therefore, this dataset is a part of design procedure. Two other datasets: TON\_IoT20 and NSL-KDD are used to evaluate whether the proposed architectures are only useful on a particular dataset or they have the  generalization capability. The hardware and software configurations that is used for designing and evaluating are as follows:
\begin{itemize}
	\item {Intel Corei7 CPU @ 3.5GHz, 16 GB of RAM and cuda-enabled nVidia GTX860m GPU}
	\item{tensorflow v2.1, Hyperopt v0.2.4 and scikit-learn libraries on MS Windows 10}
\end{itemize}
To assess the effects of compression we setup a random forest classifier  and use both original and compressed data to learn this forest. In the following subsection we discuss the design procedure and the performance of  proposed dimension reduction scheme.

\subsection{Encoder}
First, the latent space is set to five to verify which type of neural network is best for the subsequent layers of the encoder. Three types of neural networks are applied:  convolutional neural network, conventional deep neural network and recurrent neural network. The performance evaluation is done by two criteria
\begin{enumerate}
	\item{The encoder classification accuracy(equation \ref{acc})}
	\item{Mean square error of decoder in reconstructing the original input, that is calculated by:
		\begin{equation}
		\label{mse}
		MSE = \frac{1}{N}\sum\limits_{i=1}^{N} (X_{input} - X_{Reconstructed})^2
		\end{equation}.}
\end{enumerate}

Table \ref{manual_net_choose_tab} shows the results for different types of networks. When LSTM-based layers are used, the encoder classification accuracy is 94.73\% which it is the highest classification accuracy among other types and the MSE of decoder is 0.0051 which is the lowest. Therefore, the recurrent network is chosen to construct encoder.
\begin{table*}[h!]
	\footnotesize

	\caption{Results of different neural networks as hidden layer}
	\label{manual_net_choose_tab}
	\centering
	\begin{tabular}{|l|l|c|c|}
		\hline	
		Network Type & Net. Configuration & 
		$Acc._{Encoder}(\%)$  & $Decoder\ MSE$\\
		\hline
		MLP& 5,20,1 &  58.06 & - \\
		\hline
		MLP & [5,40,35,30,25,20,15,5,1] & 90.39 &0.0157\\
		\hline
		CNN &	One Conv. layer with 64 filters & 93.22 &  0.0067 \\
		\hline
		CNN &  	Two Conv. layers with 64 filters & 91.64 & 0.0068  \\
		\hline
		LSTM &  	Two recurrent layers with 180/110 LSTM units & 94.73 & 0.0051\\
		\hline

	\end{tabular}
\end{table*}
In the next step, a list consists of variety of LSTM-based neural network configurations and its hyper-parameters is defined. This list is used as Bayesian optimization search dictionary (Table \ref{encoderdict}). The max iteration of optimization algorithm is set to 200, and the winner configuration and parameters are shown in table \ref{final_archi_results}.

\begin{table*}[h!]
	\footnotesize

	\caption{Bayes optimization dictionary}
	\label{encoderdict}
	\begin{center}
		\begin{tabular}{|l|c|c|c|}
			\hline
			variable & range / selectable values & Distribution function & variation unit\\
			\hline
			1st layer LSTM units &  
			[10:200] &Uniform & q=10\\
			\hline
			2nd layer LSTM units &
			[10:200] &Uniform &q=10\\
			\hline
			Activation functions&
			'elu, 'relu', 'tanh', 'sigmoid' &Uniform & - \\
			\hline
			Epochs&
			[100,1000] &Uniform & q=50\\
			\hline
			Learning rate&
			[0.00001,0.01] &Normal (0.001, 0.01)& continuous\\
			\hline
			\multirow{4}{*}{Depth and node numbers of MLP}
			&[60,20] & \multirow{4}{*}{Uniform} &\multirow{4}{*}{-} \\
			&[100,70,40,10]& &\\
			&[50,40,30,20,10]& &\\
			&[100,80,60,40,20,10] & &\\
			\hline
			Latent space &
			2,3,4,5& Uniform&-\\
			\hline
		\end{tabular}
	\end{center}
\end{table*}

\begin{figure}[h!]
	\begin{subfigure}{.5\textwidth}
		\centering
		\includegraphics[width=1\linewidth]{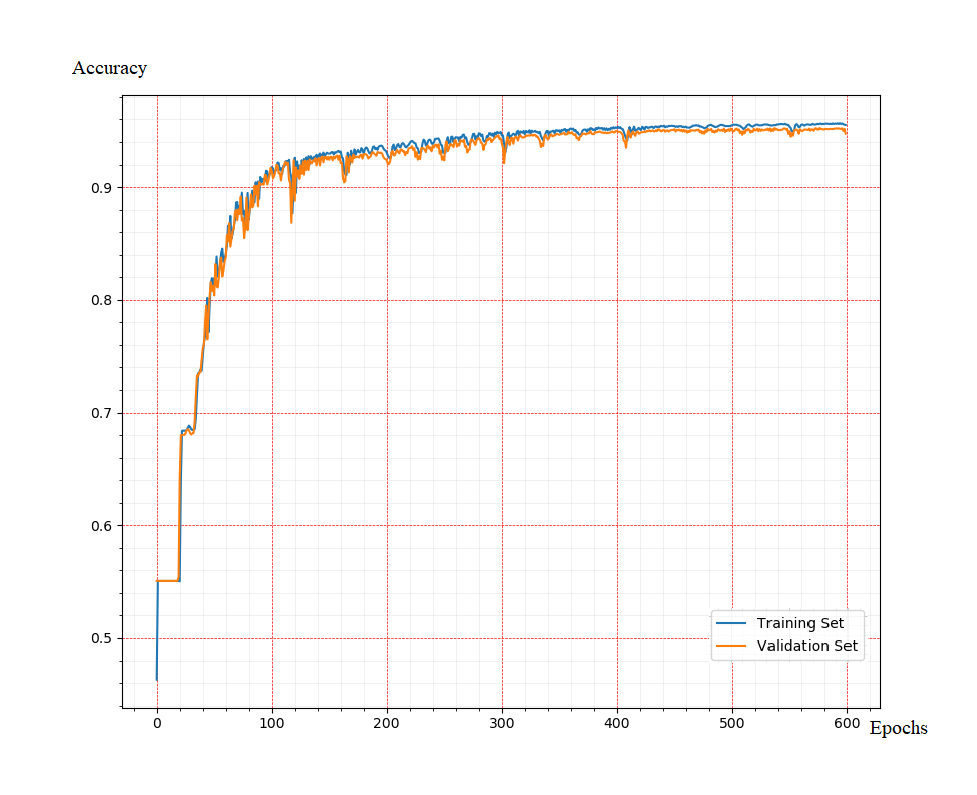}  
		\caption{Classification accuracy w.r.t epoch}
		\label{best_configa}
	\end{subfigure}
	\begin{subfigure}{.5\textwidth}
		\centering
		\includegraphics[width=1\linewidth]{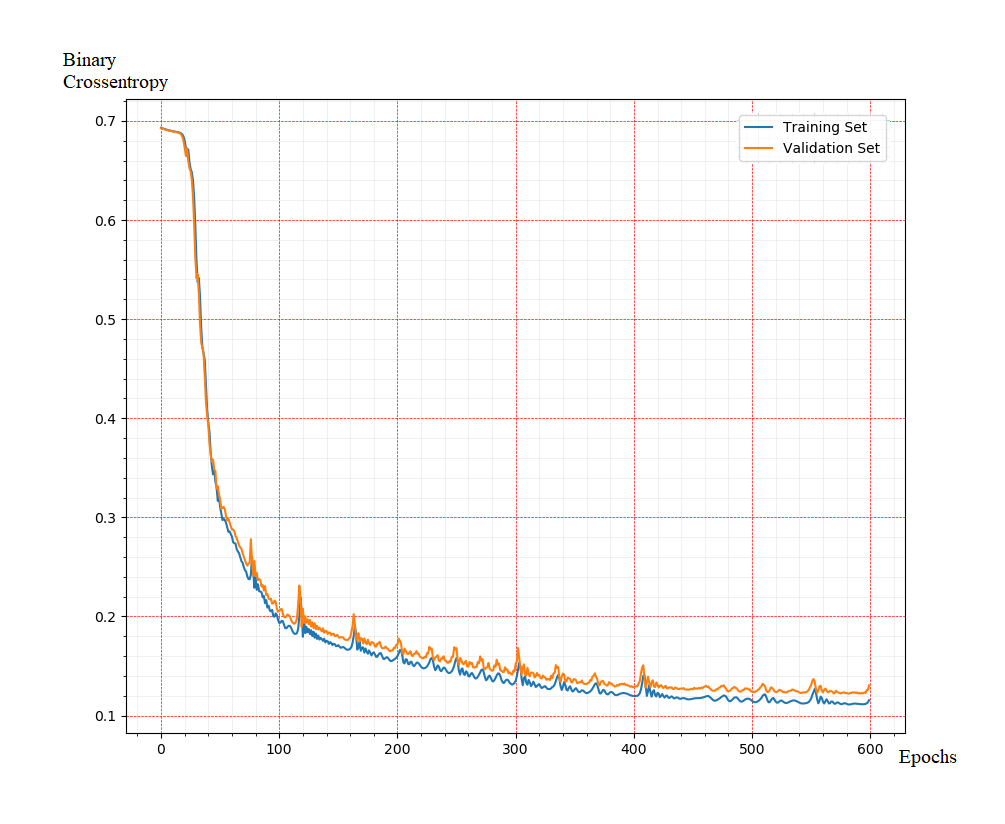}  
		\caption{Loss function w.r.t epoch }
		\label{best_configl}
	\end{subfigure}
	\caption{Classification accuracy and loss for winner configuration (LS=4)	}
	\label{best_config}
\end{figure}

\begin{table}
	\footnotesize

	\caption{Winner Conf. of Bayes method}
	\label{final_archi_results}
	\begin{center}
		\begin{tabular}{|l|c|}
			\hline
			\# of units in 1st layer&
			180\\ \hline
			\# of units in 2nd layer &
			110\\ \hline
			Activation func.&
			'relu'\\ \hline
			Epochs &
			600\\ \hline
			Learning rate&
			0.0092278\\ \hline
			MLP conf. & 
			[100,80,60,40,20,10] \\ \hline
			Latent space size &
			4\\ \hline
		\end{tabular}
	\end{center}
\end{table}
The detection accuracy and binary crossentropy loss for the winner configuration in figure \ref{best_configl} show that the validation set classification loss is tracking the training set's for higher number of epochs but there are some swings. Based on this, in the next step, we try to reduce the dimension to 3 ($LS = 3$) by increasing the number of epochs to 800 and applying learning rate decay:
\begin{equation}
decay = \frac{learning \ rate}{epochs}.
\end{equation}

\begin{figure*}[pt]
	\begin{subfigure}{.5\textwidth}
		\centering
		\includegraphics[width=1\linewidth]{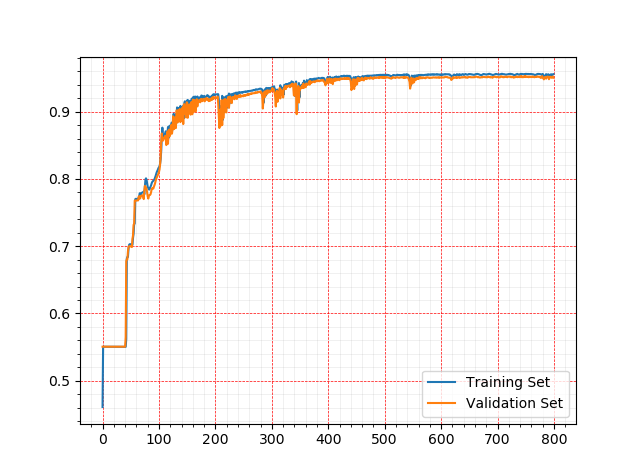}  
		\caption{Classification accuracy w.r.t epoch}
		\label{latent3A}
	\end{subfigure}
	\begin{subfigure}{.5\textwidth}
		\centering
		\includegraphics[width=1\linewidth]{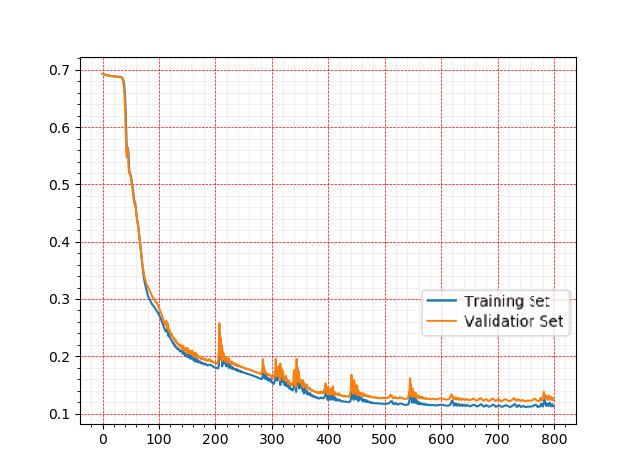}  
		\caption{Loss function w.r.t epoch}
		\label{latent3L}
	\end{subfigure}
	\caption{Classification accuracy and loss for LS=3}
	\label{latent3}
\end{figure*}

The results for detection accuracy and loss function for LS=3 (figure \ref{latent3})shows that the encoder seems to have the ability to reduce the dimension to three. To corroborate this claim, a random forest is trained with original, compressed and reconstructed datasets. Table \ref{dim_red_result_compare} shows the results of trained RF classifier on predefined UNSW-NB15 test set. 


\begin{table*}[h!]
	\footnotesize
	\caption{Random forest results on UNSW-NB15 for different input}
	\label{dim_red_result_compare}
	\begin{center}
		\begin{tabular}{|l|c|c|c|c|c|c|c|}
			\hline
			ِDataset& Val. Acc.
			& Test Acc.
			& DR &Test PR. & 
			Test FPR.   
			&Training Time (sec) & 
			Val. MSE\\
			\hline
			No Compression &
			97.68 &88.206 &84.136 &98.289 &3.121 &24.508 & -\\
			\hline
			LS = 4&  95.98 & 86.747 & 83.529& 96.532 & 6.395 & 9.318&- \\
			\hline
			LS = 3 & 95.761 & 87.584 & 83.254 &98.235 &3.188&6.537&-\\
			\hline
			Recon. of LS = 4 &
			96.004 & 87.401& 82.903& 98.323 &3.012 &80.94 &	0.0017\\
			\hline
			Recon. of LS = 3 &
			95.64 & 87.439 & 82.713& 98.607 & 2.489 & 73.584 & 0.0021\\
			\hline
			
		\end{tabular}
	\end{center}
\end{table*}

\textbf{Results Discuss:}
\begin{itemize}
	\item{Comparing the results of LS=3 and LS=4 show that in LS=3 some criteria are even improved. There is a small drop in classification accuracy and detection rate but false positive rate and precision are improved alongside with lower training time.}
	
	\item{The original data, especially in detection accuracy is doing better job.}
	\item{The decoder reconstructs the original data with lower $MSE$ when LS  = 4 }
	\item{In LS=4 the reconstructed data show improved criteria compared to others except original data}
	\item{The volume of the array which contains the over two and a half million records of UNSW-NB15 dataset is 872 MB on disk and it reduces to 29 MB when we use LS=3 and that means about 96\% compression rate.}
\end{itemize}
As it is mentioned earlier, the design and validation of the encoder and decoder is based on UNSW-NB15, therefore we evaluate the concluded architecture on two other datasets to verify the model's generalization aptitude. Tables \ref{final_res_nsl} and \ref{final_res_iot20} show the results for NSL-KDD and TON\_IoT20 datasets, respectively.

\begin{table*}[h!]
	\footnotesize

	\caption{Results of compression on NSL-KDD dataset }
	\label{final_res_nsl}
	\begin{center}
		\begin{tabular}{|l|c|c|c|c|c|c|}
			\hline
			Criterion		& Val. Acc.
			& Test Acc. 
			& DR &Test PR. & 
			Test FPR.   
			&Training Time (sec) 		\\
			\hline
			No Compression &
			99.825 & 80.127 & 67.771 & 96.195 & 3.543 &28.728 \\
			\hline
			LS = 4&  99.5& 78.104 & 63.656& 96.778 & 2.801 & 14.467 \\
			\hline
			LS = 3 & 99.555 & 78.534 & 64.724 & 96.38&3.213&9.365\\
			\hline

		\end{tabular}
	\end{center}
\end{table*}

\begin{table*}[h!]
	\footnotesize

	\caption{Results of compression on TON\_IoT20 dataset }
	\label{final_res_iot20}
	\begin{center}
		\begin{tabular}{|l|c|c|c|c|c|c|}
			\hline
			Criterion
			& Val. Acc.
			& Test Acc. 
			& DR &Test PR.& 
			Test FPR.
			&Training Time (sec) \\ 
			
			\hline
			No Compression &
			99.908 & 99.913 & 99.913 & 99.983 & 0.087 & 77.049\\
			\hline
			LS = 4&  98.884 & 98.931 &98.82 &  98.132 & 1.01& 62.363 \\
			\hline
			LS = 3 & 98.817 & 98.82 & 98.659 & 97.977 & 1.093& 43.479\\
			\hline

		\end{tabular}
	\end{center}
\end{table*}

\textbf{Results Discuss}:
\begin{itemize}
	\item{On NSL-KDD's test dataset the classification accuracy and detection rate for compressed data drop by about 2\%  and 4\%, respectively, however the classification criteria are close for LS=3 and LS=4. Therefore LS=3 can be used as a competent representative for original data.}
	\item{On TON\_IoT20 the functionality of compressed data reduces by 1\% in detection accuracy, detection rate and false alarm rate. Again, LS=3 is a quite competent representative to be used in decision engines.}
\end{itemize}

The best result of compression in reviewed papers belongs to \cite{2tier}. In \cite{2tier}, by applying 2-level compression, the final feature set size is four, beside their technique cannot be implement in a distributive manner. Our proposed scheme manage to reduce feature set size to 3 on three different datasets without any remarkable drop in classifier's functionality. In contrast to \cite{2tier,2tier_r39,tse21}, in our proposed scheme, the computation of compression can be carried out distributively. 

\section{Conclusion and Future Detection}
\label{conclusion}
In this paper, we proposed a dimension reduction scheme for IDS data using deep neural networks. The reduction procedure can be carried out in a distributive mode with a low computational overhead imposed on IoT hosts. We evaluated our proposed method on three different datasets and the results showed that the generated compressed data can be used to train classifiers and decision engines. Our proposed decoder functionality is low, hence,  rebuilding the actual input data with lower error in the central server is remained as a challenge to future works.

%

\bibliographystyle{IEEEtr}
\bibliography{refz}

\end{document}